\documentstyle[12pt,aaspp4]{article}
\newcommand{\etal}{{\it et al.} }
\newcommand{\asca}{{\it ASCA} }

\newcommand{\mcg}{MCG $-$6$-$30$-$15 }
\newcommand{\nix}{$\cdot\cdot\cdot$}

\begin{document}

\title{ON THE EVIDENCE FOR EXTREME GRAVITY EFFECTS IN  MCG $-$6$-$30$-$15}

\author{K. A. Weaver\altaffilmark{1} and T. Yaqoob\altaffilmark{2}}

\vspace{3cm}

\altaffiltext{1}{Department of Physics and Astronomy, 
Johns Hopkins University, Bloomberg Center, Baltimore, 
MD 21218. kweaver@pha.jhu.edu}
\altaffiltext{2}{NASA/ Goddard Space Flight Center, Laboratory for High
Energy Astrophysics, Greenbelt, MD 20771, USA.}
                                                                               
\begin{abstract}

We examine the unusual Fe-K$\alpha$ line profile in \mcg observed by 
$ASCA$ during a deep minimum in the source intensity.
The intense red wing and depressed blue wing of the line have 
been interpreted as evidence for extreme gravitational redshifts
in terms of emission from within six gravitational radii of a black hole.
We find that the data do not uniquely support this interpretation
and can be equally well explained by occultation of the continuum 
source and the putative line-emitting accretion disk, which we 
offer as an alternative hypothesis.  Two problems with previous 
modeling were that the equivalent width of the line during the
deep minimum was required to be unusually large ($> 1$ keV)
and the line intensity was thought to increase as the
source became dim. The occultation model does not suffer
from these problems.
Our results serve to highlight the hazards of over-interpreting
observational results which have low statistical significance, to the
extent that theoretical implications can become generally accepted
when the data do not provide a strong case for them.

\end{abstract}


\newpage

\section{Introduction}

Recently, Iwasawa \etal (1996; hereafter I96) found an extremely broad 
iron K$\alpha$ emission line with a particularly prominent red wing, extending
down to $\sim 4$ keV, during an extended observation with the
{\it Advanced Satellite for Cosmology and Astrophysics} 
({\it ASCA}; see Tanaka, Inoue \& Holt 1994).
The peculiar line profile, measured during what has been dubbed a
`deep minimum' (or DM) in the X-ray light curve, had much more 
emission in the red wing, and much less around the line rest energy
($\sim 6.4$ keV), compared to the Fe K$\alpha$ profiles thus far measured for
MCG $-$6$-$30$-$15 and other AGNs (e.g. Tanaka \etal 1995, Yaqoob \etal 1995, 
Nandra \etal 1997). Also, the equivalent width of the line was unusually
large, of the order of $\sim 1$ keV, a factor of $\sim 3$ larger than
most such measurements of broad iron K lines in Seyfert 1 galaxies. 
For the DM state of MCG $-$6$-$30$-$15, I96 rejected the standard model 
for the Fe-K$\alpha$ line profile (e.g. see Fabian \etal 1989), in which 
the line is emitted in an accretion disk rotating about a Schwarzschild 
black hole, with the inner disk extending no closer than the radius
for marginally stable orbits, (i.e. 6 gravitational radii,
or $6 r_{g}$). Instead, the extreme gravitational redshifts implied
by the bloated red wing of the line were explained either in terms of
emission from inside the last stable orbit around a Schwarzschild hole 
(Reynolds \& Begelman 1997) or emission in a disk rotating about a Kerr 
black hole (e.g. I96; Dabrowski \etal 1997; Bromley, Miller \& Pariev
1998).  In the latter case the last stable 
orbit extends down to $1.24r_{g}$ for a maximally rotating Kerr 
black hole. In both cases, the red wing of the line is accounted
for by emission closer to the black hole event horizon, so the photons
can experience the effects of very strong gravity, resulting in
extremely large redshifts.   

We point out that in terms of fitting the 
DM data for \mcg with the standard model
(Schwarzschild hole, disk extending to $6 r_{g}$) and the
Kerr model (disk extending to $1.24 r_{g}$),
the largest difference
in the fitting statistic, $\Delta \chi^{2}$, is 6.2 for the same
number of free model parameters (see rows 1 and 3
in Table 3 of I96 who assume a disk inclination
of 30$^{\circ}$ and an outer radius of $15.5r_{g}$).
While this may be interpreted as being formally statistically significant,
\asca spectral fits do not in general (and in this case, in particular)
include the effects of systematic errors which could reduce the 
overall significance of the result. Since the implications of really
being able to observe X-rays inside of $6 r_{g}$ and even closer
to a black hole event horizon are so far reaching (e.g. see Fabian 1997) 
it is important to investigate the robustness of the result for
MCG $-$6$-$30$-$15, the only case thus far reported.

\section{ASCA Data}

\asca observed \mcg for over 4 days starting 1994, July 23.
\asca has four identical, thin-foil, light-weight X-ray
telescopes (XRT) which focus X-rays onto one of two
Solid-state Imaging Spectrometers (SIS) or one of two
Gas Imaging Spectrometers (GIS, see Ohashi \etal 1991).
See Tanaka \etal (1994) for a summary of the
\asca mission and focal-plane detectors.
The SIS sensors, each one consisting of four CCD (charge coupled
device) chips were operated in a mode in which only one chip was
exposed (1-CCD mode) and the data
were accumulated in FAINT mode.
Hereafter
the two SIS sensors are referred to as SIS0 and SIS1 and the two GIS sensors
as GIS2 and GIS3.
The data reduction and selection criteria are similar 
to those described in Yaqoob \etal (1994). 

The lightcurve of the entire observation has been presented elsewhere
(I96; Reynolds \etal 1997; Yaqoob \etal 1997). 
We use exactly
the same time intervals defined by I96 to extract spectra of 
the DM state (interval $i-7$ in their Figure 2) and 
the flare state (interval $i-3$), as well as the   
average (total) spectrum.
For the DM, we obtained $3-10$ keV count rates in the range 
0.13 to 0.16 counts s$^{-1}$
and exposure times in the range 13.2 to 13.3 ks for the four instruments.
Figure 1 shows the ratio of the data in the DM 
to the best-fitting power-law model
($\Gamma$ = 1.92, $N_{\rm H}$ = $1.1\times10^{22}$ cm$^{-2}$)
using data only in the energy ranges $2-3$ keV and $8-10$ keV.
The excess above the underlying power-law is due to the Fe-K$\alpha$
line emission.  As pointed out by
I96, the emission on the blue side of the line is
unusually diminished compared to the red side.
The portion of the lightcurve 
containing the DM state is shown in Figure 2. 

\section{Standard Disk-line Spectral Fits}

Using data between 3 and 10 keV from all four instruments,
we fitted the Fe-K$\alpha$ line for the average, flare and DM spectra
with our `baseline' model in which the line photons are
emitted in a disk rotating around a central Schwarzschild black hole
(e.g., Fabian \etal 1989).  The parameters are
$\theta_{\rm obs}$ (inclination angle of the disk normal relative to
the observer), $r_{i}$ (inner disk radius),
$r_{o}$ (outer disk radius), $q$ (power-law index characterizing the
line emissivity as $\propto r^{-q}$),
$I_{\rm D}$ (line intensity), and $E_{D}$ (line energy in the
disk rest frame).  
The inner radius, $r_{i}$, was fixed at 
$6 r_{g}$ where $r_{g} \equiv GM/c^{2} $ (i.e. 
the last stable orbit). There
is interplay between $E_{D}$ and the other line parameters,
so $E_{D}$ was fixed at 6.4 keV in the rest frame, 
corresponding to fluorescence in a cold disk.
The results are shown in Table 1, models SH1, 
SH2, and SH3.

We repeated the above fitting,   
but this time replacing the
Schwarzschild black hole with a maximally rotating Kerr black hole
(see e.g., Laor 1991).
Now the inner radius is fixed at $1.24 r_{g}$, the minimum
value, or last stable orbit for the Kerr metric.
The spectral fit was repeated with $E_{D}$ fixed at
6.7 keV (corresponding to He-like Fe).
For the average and flare spectra, a Kerr profile does
not improve the fit, but for the DM,
the Kerr model provides a statistically better fit than
the Schwarzschild model ($\Delta \chi^{2} \sim 5$).
The results for the DM are shown 
in Table 1, models KR1 and KR2.
The basic reason why the Fe-K$\alpha$ line during the DM
is better described by
emission from matter closer than $6 r_{g}$ 
is that the observed strength of the blue peak of the line
is too small for the observed strength of the red wing.
This is compensated by the extra gravitational redshifts obtained
by placing the emitter closer to the black hole, increasing the 
intensity of the 
red wing relative to the blue.
Our results confirm those of I96, although  
the best-fitting parameters differ slightly  
because we fix $q$ at 2.5 and allow $\theta_{\rm obs}$ and 
$r_{o}$ to be free, the contrary to their technique.  

\section{Spectral Fits with an Occultation Model}

We tested whether we could fit the DM Fe-K$\alpha$ line profile
without placing line-emitting matter closer to the black hole than
$6r_{g}$ by obscuring part of the standard Schwarzschild disk
with optically-thick matter.
Figure 3 illustrates how different parts of a Schwarzschild disk
affect the Doppler and gravitational shift of line photons.
Defining $g=E({\rm observed})/E({\rm emitted})$ (including both Doppler
and gravitational effects), the shaded part labeled blue
corresponds to $g>1$ and the shaded part labeled red corresponds to
$g<(1/1.2)$. In our simple-minded model,
the occultation of the disk is in the form 
of a strip running along the $x$ or $y$ directions, with half-width
$y_{0}$ or $x_{0}$
respectively, passing symmetrically over
the center of the disk.
Thus, line emission from
either  $|y|<y_{0}$ or $|x|<x_{0}$ is blocked from view and
the predicted line profiles for the two cases are shown in 
Figures 4a and 4b respectively (for several values of the half-width).
We use this model as a hypothesis, since we do not know the  
geometry of the optically-thick clouds
(there are plenty of arguments for their existence, e.g. Guilbert
\& Rees 1988). The clouds could be spherical, filament-shaped, or
neither.
Our simple model in some sense mimics an optically-thick cloud,
whatever its shape, moving
over the disk, resulting in some kind of average line profile.
Clearly, the profiles in Figure 4a, in which 
much of the blueshifted emission is obscured, are relevant for the DM data.

For the spectral fits with the obscuration model, we  
fix $E_{D} =6.4$ keV and $r_{i}=6r_{g}$ as before,
and $r_{0}=18.5 r_{g}$ from the Schwarzschild 
fit (Table 1, SH3).  The best-fitting parameters 
are shown in Table 1, model OC1.
This fit is marginally 
better than the Kerr model with $E=6.4$ keV (model KR1) and marginally
worse than the Kerr model with $E=6.7$ keV (model KR2) and so 
we conclude that the occulted, Schwarzschild disk-line 
model describes the spectrum during the DM 
just as well as the standard Kerr model.

The above fits include only Galactic absorption
because we assume that there is no significant 
absorption from the `warm absorber' present 
above 3 keV that will affect the 
shape of the continuum 
(Fabian \etal 1994; Reynolds \etal 1995; Otani \etal 1996).
This was also the
assumption made by I96 and is based on observations 
of the warm absorber in MCG$-$6-30-15 
in higher flux states.
However, if absorption were present, this 
could also affect measurements of the Fe-K$\alpha$ line. 
Figure 5 shows the 
data/model ratio when the DM spectrum is fitted
with Galactic absorption and a power law with the best-fitting 
photon index of the average spectrum ($\Gamma = 1.98$).  
The deficiency of low-energy photons due to the ionized 
absorber is significant, even {\it above} 3 keV, and 
implies that the flattening of the continuum in the DM 
(Table 1, models SH3, KR1, KR2, OC1) is due to 
absorption.  Thus, the {\it intrinsic} spectral slope
during the DM is not required to change.

We therefore fitted the DM spectrum again with 
the occultation-line model with $\Gamma$ fixed at 
the average value of $\Gamma = 1.98$, but with    
free absorption (Table 1, model OC2). The 
number of free parameters
is the same as before, but
$\Delta\chi^2 = 5.7$ compared to the
best-fitting Kerr model (model KR2).
The occultation description of the DM
is thus as viable as the  
scenarios posited by 
I96, Reynolds \& Begelman (1997), Dabrowski \etal (1997),
and Bromley \etal (1998).

\section{Discussion and Conclusions}

We have examined the Fe-K$\alpha$ line
profile during the deep minimum (DM) flux state 
for MCG$-$6-30-15 with an `occultation model'
in which the continuum source and much of the blueshifted part
of the emission from the disk is obscured by an 
optically-thick cloud.  This model describes the observed line
profile as well as the Kerr black-hole/disk model, but does not 
require emission from inside $6r_g$ (the last stable 
orbit for a Schwarzschild black hole).  One difficulty with 
the Kerr interpretation proposed by I96 and others 
is that it doesn't explain why Kerr effects should only 
manifest themselves in the DM.  The occultation model 
does not suffer from this problem.  Also,
the line intensity is {\it not} required
to increase relative to the average line intensity,
as has been previously claimed.  Nor is the equivalent
width (EW) of the Fe-K line during the DM required to be unusually
large. We obtain $\rm EW \sim 400$ eV for the DM {\it and} flare-states,
and $\rm EW \sim 470$ eV for the average line.  

In addition to the optically-thick obscurer that blocks 
most of the continuum region, the data above 
3 keV are affected by the O VIII edge 
of the warm absorber that becomes stronger 
in the low-flux state (Otani \etal 1996).  We  
measure an additional column density of $\sim 2 \times 10^{22}$
cm$^{-2}$ in the DM, which is  
a factor of $\sim 2$ increase compared to 
the high flux state (Fabian \etal 1994).  
The increase in column density is consistent 
with a picture in which the material that is photoionized by the 
central source becomes less opaque to soft X-rays 
due to a reduction in ionization when the continuum 
is blocked.  Note that the extra absorption we directly measure
is {\it not} necessarily related to the optically-thick clouds
responsible for obscuring the Fe-K$\alpha$ line emission. 

To conclude, 
we have shown that the unusual Fe-K$\alpha$ line profile in \mcg 
during a deep minimum in intensity does not necessarily require 
line emission from within $6r_{g}$ and that the data are 
not adequate enough to distinguish between the Schwarzschild and  
Kerr solutions.  Sulentic \etal (1998) 
have also made the suggestion that a Kerr black hole
is not required, proposing that the
true shape of the Fe-K$\alpha$ line is a double Gaussian,
but offer no physical model.
We have shown that the deep-minimum line profile can be
explained by occultation of a standard Schwarzschild disk.
This model is consistent with the detailed temporal
profile of the continuum intensity dip. 
Our results serve to highlight the hazards of over-interpreting
observational results which have low statistical significance, to the
extent that theoretical implications can become generally accepted
when the data do not provide a strong case for them.

\clearpage
\newpage

\begin{deluxetable}{llccccccc}
\tablenum{1}
\small              
\tablewidth{0pt}
\tablecaption{MCG$-$6-30-15: Relativistic Disk-Model Spectral 
Fits ($3-10$ keV) for the Fe K$\alpha$ line$^{1}$.
\label{tab:jointfits}}
\tablecolumns{9}
\tablehead{
\colhead{Data set} & \colhead{Model$^2$} & \colhead{$\Gamma$} &
  \colhead{$r_{o}/r_{g}$} & \colhead{$\theta_{\rm obs}$} & 
  \colhead{$I_{\rm Fe}$ $^{3}$} & \colhead{EW} & \colhead{$y_{0}/r_{g}$} & 
  \colhead{$\chi^2/\nu$} \nl
\colhead{} & \colhead{} & \colhead{} & \colhead{} &
  \colhead{(degrees)} & \colhead{} & \colhead{(eV)} & \colhead{} &
  \colhead{}
}
\startdata
Average & SH1 & 1.98$^{+0.04}_{-0.04}$ & 15.6$^{+5.7}_{-4.4}$ 
   & $33.7^{+1.8}_{-1.9}$ & 1.88$^{+0.30}_{-0.31}$ 
   & $473 ^{+75}_{-78}$ & \nix & 1417.9/1355 \nl
Flare & SH2 & 2.00$^{+0.07}_{-0.06}$ & 6.1$^{+2.0}_{-0.01 \dagger}$ 
   & 39.4$^{+2.4}_{-6.5}$ & 2.29$^{+0.78}_{-0.80}$
   & 410$^{+142}_{-140}$ & \nix & 925.6/899 \nl
Deep Min. & SH3 & 1.71$^{+0.17}_{-0.17}$ & 18.5$^{+19.4}_{-9.3}$
  & $27^{+10}_{-10}$ & 1.52$^{+0.74}_{-0.68}$ & $665 ^{+324}_{-298}$
  & \nix & 305.2/323 \nl
Deep Min. & KR1 & 1.73$^{+0.19}_{-0.17}$ & 19.8$^{+30.0}_{-18.56 \dagger}$ 
  & 31$^{+39}_{-12}$ & $2.28^{+1.04}_{-0.95}$ & 1070$^{+488}_{-446}$ &
  \nix & 299.6/323 \nl
Deep Min. & KR2$^{4}$ & 1.73$^{+0.21}_{-0.18}$ & $13.3^{+15.4}_{-12.06 \dagger}$ 
  & 27$^{+39}_{-10}$ & 2.42$^{+1.22}_{-1.04}$ & 1130$^{+570}_{-486}$ 
  & \nix & 297.6/323 \nl
Deep Min. & OC1$^{5}$ & 1.74$^{+0.20}_{-0.17}$ & 18.5f &
  46$^{+43}_{-23}$ & 1.63$^{+1.07}_{-0.71}$ & $481^{+316}_{-210}$ 
  & $8.0^{+4.1}_{-8.0}$  &  298.2/323 \nl
Deep Min. & OC2$^{5,6}$ & 1.98f & 18.5f & $51^{+38}_{-24}$
   & $1.53^{+0.78}_{-0.76}$ & 404$^{+206}_{-201}$ &
   $8.8^{+4.1}_{-7.2}$ & 291.9/323 \nl
\tablenotetext{1}{Errors are 90\% confidence for 4 interesting parameters
($\chi^2$ + 7.78). Fixed parameters are denoted by `f'.
The center energy of the iron line in the disk frame
is fixed at 6.4 keV unless otherwise stated. $N_{\rm H}$(Gal.) 
= $5\times10^{20}\ \rm cm^{-2}$ is assumed for all fits.}
\tablenotetext{2}{Theoretical models for the 
iron line are SH: Schwarzschild profile, KR: Kerr profile,
OC: occulted, Schwarzschild disk-line model.} 
\tablenotetext{3}{Line intensity in units $10^{-4} \rm \ photons 
\ cm^{-2} \ s^{-1}$. Intensities and equivalent
widths (EW) are in the {\it observed} frame.}
\tablenotetext{4}{The iron line center energy in this Kerr model is fixed
at 6.7 keV in the disk frame.}
\tablenotetext{5}{The occulted disk model is described in the text. The
parameter $y_{0}$ is the half-width of an obscuring strip running
symmetrically over the center of the disk (see Figure 3). }
\tablenotetext{6}{In this model there is an
additional absorbing column with $N_{H}
=1.96^{+1.29}_{-1.31} \times 10^{22} \rm \ cm^{-2}$.}
\tablenotetext{\dagger}{The lower bounds on $r_{o}$ for these fits are
not true statistical errors but represent the radius of last stable orbit around
a maximally rotating Kerr black hole ($1.235r_{g}$).}
\enddata
\end{deluxetable}

\clearpage
\newpage

\begin{figure}
\plotfiddle{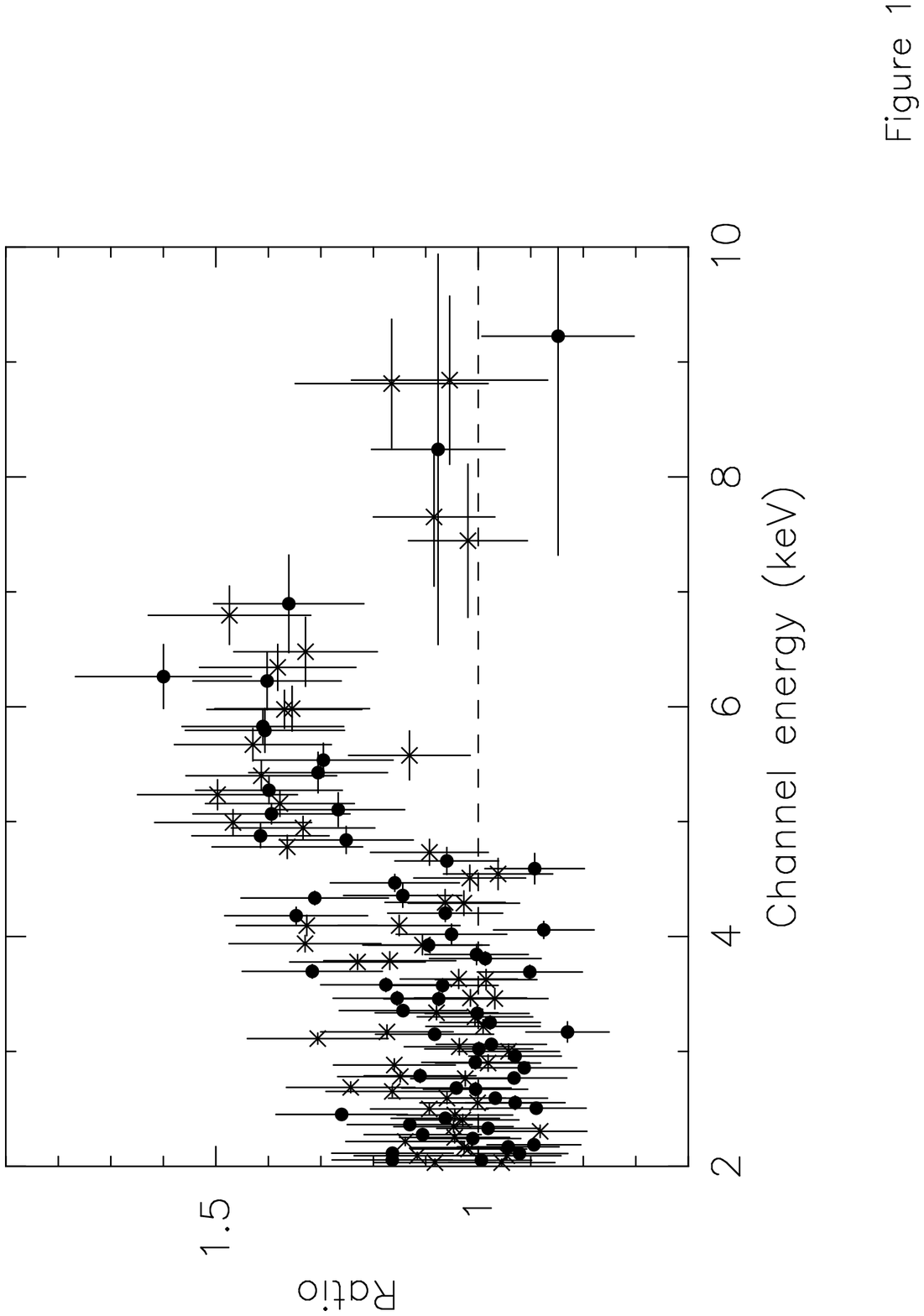}{150pt}{-90}{75}{75}{-325}{395}
\figcaption[ ]{Ratio of the $ASCA$ data (all detectors)
from the deep minimum to a continuum
model that consists of the best-fitting absorbed power law.
The data between 3 and 8 keV were ignored for the fit.
Filled circles correspond to
SIS0 and SIS1 data and crosses correspond to GIS2 and GIS3 data.
}
\end{figure}

\clearpage
   
\begin{figure}
\plotfiddle{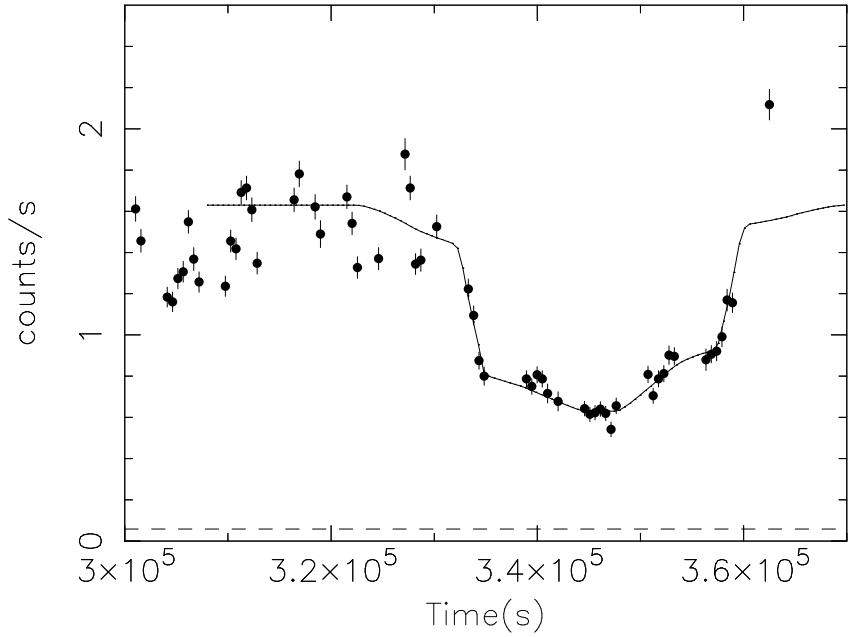}{150pt}{-90}{90}{90}{-367}{400}
\figcaption[ ]{The deep minimum (DM) light curve with the
model from McKernan and Yaqoob (1998) in which the
temporal profile of the DM lightcurve can
be explained by an obscuring body passing over the
disk and central source. To explain the temporal data,
the part of the disk for which the emission is blueshifted must
be occulted first, and it remains occulted
for a large part of the duration of the intensity dip.
The dashed line is the mean background level.
}
\end{figure}

\clearpage
     
\begin{figure}
\plotfiddle{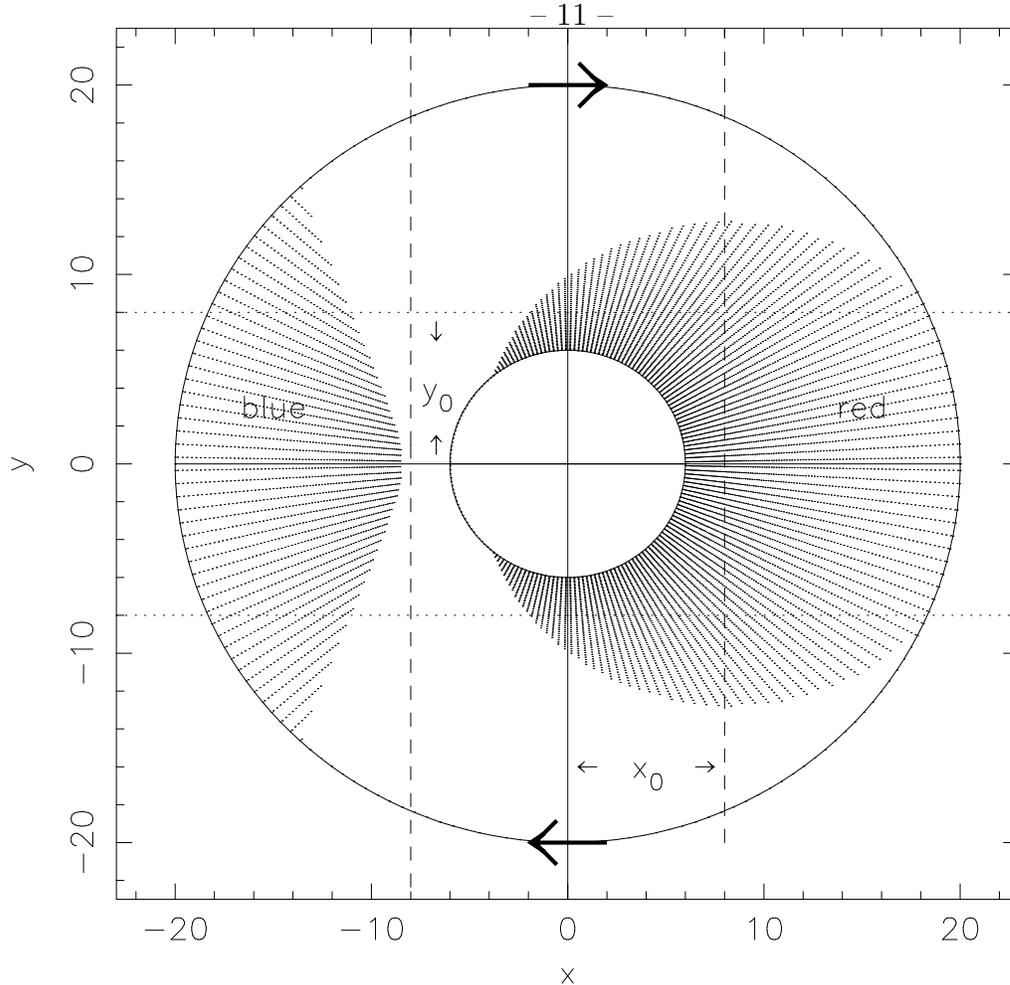}{150pt}{-90}{90}{90}{-358}{445}
\figcaption[ ]{Illustration of
how different parts of the Schwarzschild disk
affect the Doppler and gravitational shift of line photons.
Defining $g=E({\rm observed})/E({\rm emitted})$ (including both Doppler
and gravitational effects), the shaded part labeled blue
corresponds to $g>1$ and the shaded part labeled red corresponds to
$g<(1/1.2)$. The obscuration model is such that lines-of-sight to the
region inside $|x|<x_{0}$ or $|y|<y_{0}$ are blocked from view (see text).
In this
diagram the disk is inclined at 30$^{\circ}$, such that the disk
normal is pointing downwards towards the
negative $y$ axis. The inner and outer disk radii are
$r_{i}=6r_{g}$ and $r_{o}=20r_{g}$.
}
\end{figure}

\clearpage

\begin{figure}
\plotfiddle{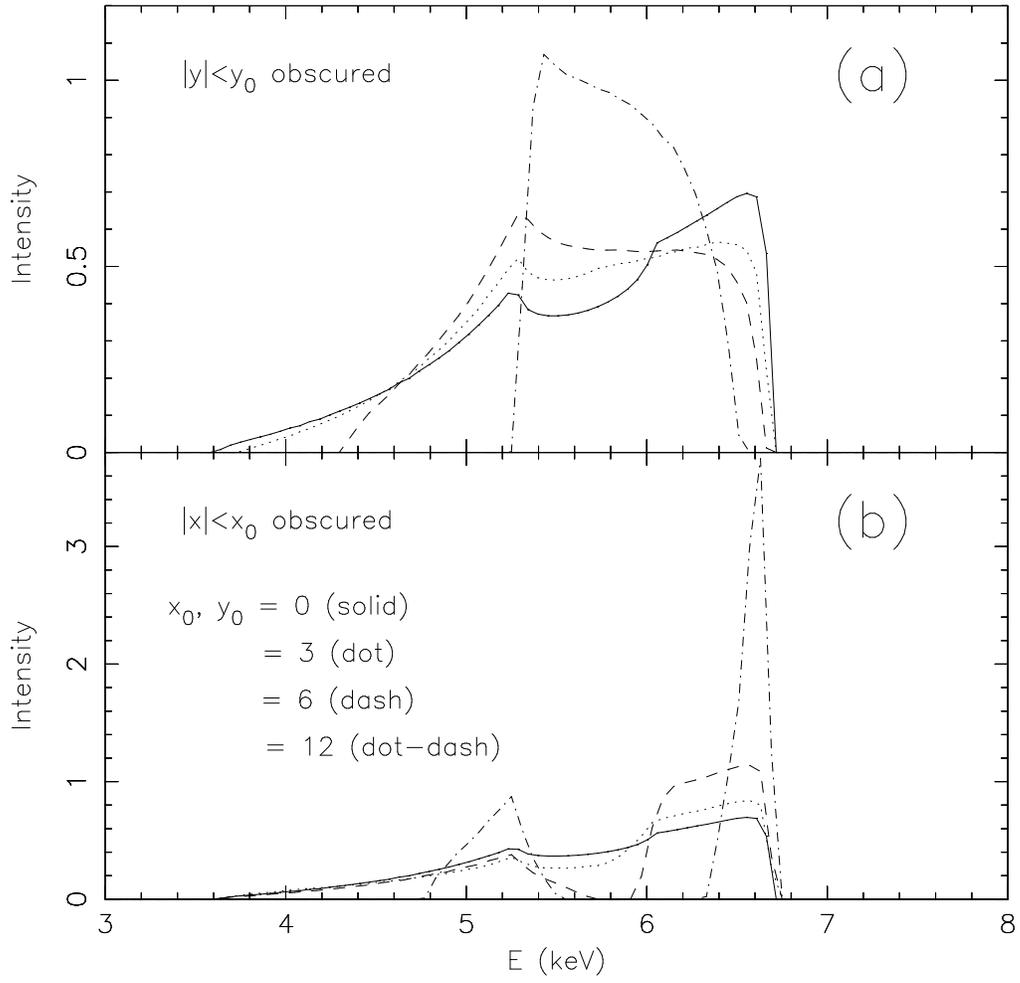}{150pt}{-90}{75}{75}{-293}{420}
\figcaption[ ]{Theoretical calculations of the iron line profile
from an obscured Schwarzschild disk with the dimensions of the
obscurer ($x_{0}$ and $y_{0}$) as shown and the remaining parameters as in
Figure 3.
}
\end{figure}

\clearpage
 
\begin{figure}
\plotfiddle{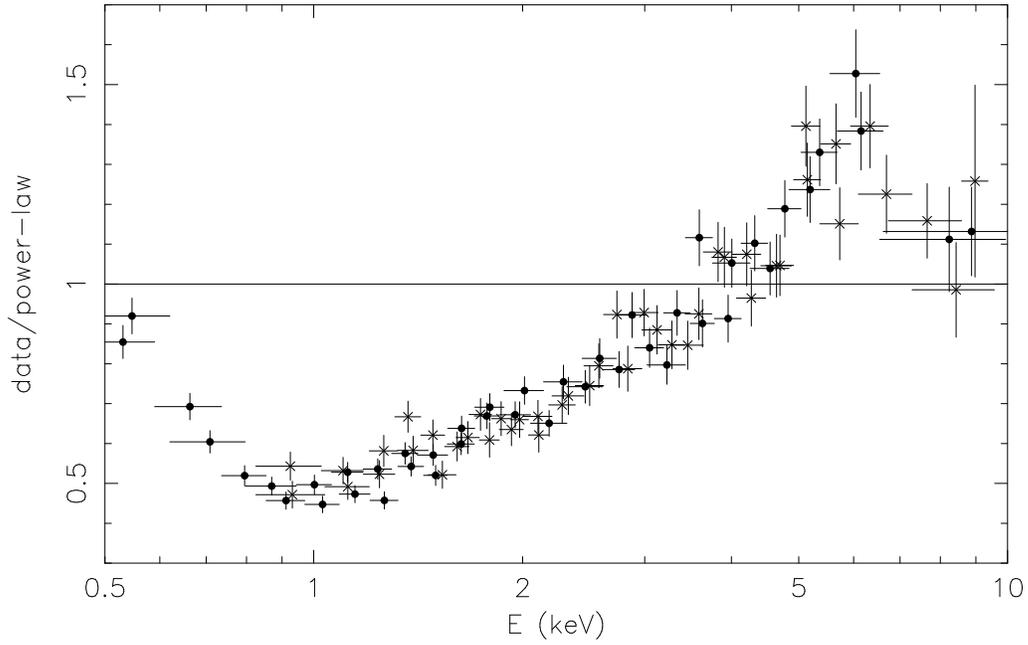}{150pt}{-90}{75}{75}{-295}{320}
\figcaption[ ]{Ratio of the data to model when the deep minimum
spectra (from the four $ASCA$ instruments) are fitted with only
Galactic absorption and a power law with the slope fixed at the
value obtained from the average 4.2-day spectrum (i.e., $\Gamma=1.98$,
see model SH1 in Table 1). The extra absorption during the
deep minimum is clearly evident, even above 3 keV.
Filled circles correspond to
SIS0 and SIS1 data and crosses correspond to GIS2 and GIS3 data.
}
\end{figure}
  
\end{document}